\documentclass{article}%
\usepackage{amsfonts}
\usepackage{amsmath}
\usepackage{amssymb}
\usepackage{graphicx}
\usepackage{hyperref}%
\providecommand{\U}[1]{\protect\rule{.1in}{.1in}}
\begin{document}

\title{On Rotations as Spin Matrix Polynomials}
\author{T. L. Curtright$^{\S }$ and T. S. Van Kortryk\smallskip\\{\small curtright@miami.edu and vankortryk@gmail.com\medskip}\\$^{\S }$Department of Physics, University of Miami\\Coral Gables, FL 33124-8046, USA}
\date{}
\maketitle

\begin{abstract}
Recent results for rotations expressed as polynomials of spin matrices are
derived here by elementary differential equation methods. \ Structural
features of the results are then examined in the framework of biorthogonal
systems, to obtain an alternate derivation. \ The central factorial numbers
play key roles in both derivations.

\end{abstract}

\section*{Introduction}

Curtright, Fairlie, and Zachos (CFZ) recently obtained explicit and intuitive
results \cite{CFZ} expressing the rotation matrix for \emph{any} quantized
angular momentum $j$ as a polynomial of order $2j$ in the corresponding
$\left(  2j+1\right)  \times\left(  2j+1\right)  $ spin matrices
$\boldsymbol{\hat{n}\cdot J}$ that generate rotations about
axis\ $\boldsymbol{\hat{n}}$. \ While many previous studies of this or closely
related problems can be found in the literature --- beginning with the work of
Wigner in the 1930s \cite{Wigner,WignerAgain} and then, after a lengthly
hiatus, continuing in the 1960s and subsequently with direct attacks on the
problem by Lehrer-Ilamed \cite{Illamed}, van Wageningen \cite{vanW}, and
others \cite{W}-\cite{T} --- none of these other studies succeeded to find
such simple, compact expressions for the coefficients in the spin matrix
polynomial, as elementary functions of the rotation angle, as those obtained
by CFZ. \ For each angle-dependent coefficient in the polynomial, the explicit
formula found by CFZ involves nothing more complicated than a truncated series
expansion for a power of the $\arcsin$ function.

The CFZ formula for a rotation through an angle $\theta$ about an axis
$\boldsymbol{\hat{n}}$, valid for any spin $j\in\left\{  0,\frac{1}{2}%
,1,\frac{3}{2},\cdots\right\}  $, is
\begin{equation}
\exp\left(  i~\theta~\boldsymbol{\hat{n}\cdot J}\right)  =\sum_{k=0}^{2j}%
\frac{1}{k!}\left.  A_{k}^{\left[  j\right]  }\left(  \theta\right)  \right.
\left(  2i~\boldsymbol{\hat{n}\cdot J}\right)  ^{k}~, \label{the result}%
\end{equation}
where the angle-dependent coefficients of the various spin matrix powers are
given simply by%
\begin{equation}
A_{k}^{\left[  j\right]  }\left(  \theta\right)  =\sin^{k}\left(
\theta/2\right)  ~\left(  \cos\left(  \theta/2\right)  \right)  ^{\epsilon
\left(  j,k\right)  }~\operatorname*{Trunc}_{\left\lfloor j-k/2\right\rfloor
}\left[  \frac{1}{(\sqrt{1-x})^{\epsilon\left(  j,k\right)  }}\left(
\frac{\arcsin\sqrt{x}}{\sqrt{x}}\right)  ^{k}\right]  _{x=\sin^{2}\left(
\theta/2\right)  }\ . \label{the coefficients}%
\end{equation}
Here, $\left\lfloor \cdots\right\rfloor $ is the integer-valued floor
function\footnote{For any $r\in%
\mathbb{R}
$, $\left\lfloor r\right\rfloor =\sup\left\{  n\in%
\mathbb{Z}
\text{ such that }n\leq r\right\}  $.} and $\operatorname*{Trunc}%
\limits_{n}\left[  f\left(  x\right)  \right]  $ is the $n$th-order Taylor
polynomial truncation for any $f\left(  x\right)  $ admitting a power series
representation:%
\begin{equation}
f\left(  x\right)  =\sum_{m=0}^{\infty}f_{m}x^{m}%
\ ,\ \ \ \operatorname*{Trunc}_{n}\left[  f\left(  x\right)  \right]
\equiv\sum_{m=0}^{n}f_{m}x^{m}\ .
\end{equation}
In addition, $\epsilon\left(  j,k\right)  $ is a binary-valued
function\footnote{For $k$ integer and $j$ either integer or semi-integer,
$\epsilon\left(  j,k\right)  =\left(  1-\left(  -1\right)  ^{2j-k}\right)
/2$.} of $2j-k$ that distinguishes even and odd integers: $\ \epsilon\left(
j,k\right)  =0$ for even $2j-k$, and $\epsilon\left(  j,k\right)  =1$ for odd
$2j-k$. \ 

As observed in \cite{CFZ}, the results (\ref{the coefficients}) display the
limit $j\rightarrow\infty$ for fixed $k$ in a beautifully intuitive way. \ In
that limit, the truncation is lifted to obtain trigonometrical series for the
\emph{periodicized} $\theta^{k}$ monomials. \ But even as $j\rightarrow\infty
$, integer $j$ (bosonic) and semi-integer $j$ (fermionic) coefficients are
clearly distinguished by a relative sign flip for $\theta\in\left[  \pi
,3\pi\right]  \operatorname{mod}\left(  4\pi\right)  $. \ This is evident upon
plotting the first few coefficients for very large spins. \ Following
\cite{CFZ}, a few examples are shown in Appendix G.

In practice, for finite $j$ of reasonable size, the truncations needed to
evaluate (\ref{the coefficients}) are easily obtained as a matter of course by
machine computation, for example by using either \emph{Maple} or
\emph{Mathematica}. \ Nevertheless, it is interesting and relevant for the
analysis to follow that Taylor series for powers of cyclometric functions\ can
be expressed in terms of $t\left(  m,n\right)  $, the so-called \emph{central
factorial numbers of the first kind} \cite{Riordan,CFN}. \ Thus for
$\left\vert z\right\vert \leq1$ and non-negative integer $n$ (cf. Theorem
(4.1.2) in \cite{CFN}), \
\begin{equation}
\left(  \arcsin\left(  z\right)  \right)  ^{n}=\frac{n!}{2^{n}}\sum
_{m=n}^{\infty}\frac{\left\vert t\left(  m,n\right)  \right\vert }{m!}\left(
2z\right)  ^{m}\ . \label{cyclometric power series}%
\end{equation}
Note that the coefficients in these Taylor series are all non-negative.
\ Also, $t\left(  m,n\right)  =0$ for odd $m+n$, so the expansions
(\ref{cyclometric power series}) for even (odd) powers of $\arcsin\left(
z\right)  $ are indeed even (odd) functions of $z$. \ In general, the values
of $t\left(  m,n\right)  $\ are defined by and obtained from simple
polynomials, as described in Appendix A.

Incorporating (\ref{cyclometric power series}) into the expression for the
coefficients (\ref{the coefficients}) gives
\begin{equation}
A_{k}^{\left[  j\right]  }\left(  \theta\right)  =\frac{k!}{2^{k}}\sum
_{m=k}^{2j}\frac{2^{m}}{m!}\left\vert t\left(  m,k\right)  \right\vert
\sin^{m}\left(  \theta/2\right)  \text{ \ \ for even\ }2j-k\ .
\label{the coefficients as cfns}%
\end{equation}
As firmly established in \cite{vanW,CFZ}, the remaining coefficients in
(\ref{the result}) may then be obtained from
\begin{equation}
A_{k-1}^{\left[  j\right]  }\left(  \theta\right)  =\frac{2}{k}\frac
{d}{d\theta}~A_{k}^{\left[  j\right]  }\left(  \theta\right)  \text{ \ \ for
odd\ }2j-k+1\ . \label{oddbydiff}%
\end{equation}

\section*{A Derivation Using Differential Equations}

The goal here is to derive (\ref{the coefficients as cfns}), hence to
establish (\ref{the coefficients}) and (\ref{the result}), by using elementary
results extant in the literature, and by using a simple Lemma, namely,%
\begin{equation}
\left(  2\boldsymbol{\hat{n}\cdot J}\right)  ^{2j+1}=-\sum_{m=0}%
^{2j}2^{1+2j-m}\times t\left(  2+2j,1+m\right)  \times\left(
2\boldsymbol{\hat{n}\cdot J}\right)  ^{m}\ . \label{Lemma}%
\end{equation}
This Lemma is established in Appendix B. \ Given the Lemma and well-known
properties of the central factorial numbers, a proof of
(\ref{the coefficients as cfns}) follows directly. \ To see this, begin by
considering integer values of $j$. \ 

For integer $j$ and even index coefficients, the results
\begin{equation}
A_{2k}^{\left[  j\right]  }\left(  \theta\right)  =\frac{\left(  2k\right)
!}{4^{k}}\sum_{m=2k}^{2j}\frac{2^{m}}{m!}\left\vert t\left(  m,2k\right)
\right\vert \sin^{m}\left(  \theta/2\right)  \label{IntegerSpinEvenCoef}%
\end{equation}
may be obtained by verifying that these series for various $k$ are in fact
solutions of the second-order equations
\begin{equation}
A_{2k-2}^{\left[  j\right]  }\left(  \theta\right)  =\frac{4}{2k\left(
2k-1\right)  }\frac{d^{2}}{d\theta^{2}}~A_{2k}^{\left[  j\right]  }\left(
\theta\right)  +\left(  -4\right)  ^{j-k+1}\times t\left(  2+2j,2k\right)
\times\frac{\left(  2k-2\right)  !}{\left(  2j\right)  !}~A_{2j}^{\left[
j\right]  }\left(  \theta\right)  \ , \label{2ndOrderDiffEqn}%
\end{equation}
with proper behavior near $\theta=0$. \ The correct small $\theta$ behavior
follows immediately from that of the exponential on the LHS of
(\ref{the result}), and is easily seen to hold for the series
(\ref{IntegerSpinEvenCoef}) and their first derivatives with respect to
$\theta$. \ That these second-order differential equations and initial
conditions are necessary and sufficient is a straightforward consequence of
the first derivative relations carefully derived in Section 6 of \cite{CFZ},
and of the Lemma. \ By dealing with the second-order equation
(\ref{2ndOrderDiffEqn}) instead of directly with first derivatives of the
coefficients, one can avoid the cosine and $\sqrt{1-x}$ factors in
(\ref{the coefficients}). \ The coefficients must also obey higher order
differential equations, as discussed in \cite{vanW,WW}. \ These higher order
equations are also satisfied by (\ref{IntegerSpinEvenCoef}), but it is
unnecessary to show this here.

To show that (\ref{2ndOrderDiffEqn}) is indeed satisfied by
(\ref{IntegerSpinEvenCoef}), first compute the second derivative of the
series:
\begin{equation}
\frac{4}{2k\left(  2k-1\right)  }\frac{d^{2}}{d\theta^{2}}~A_{2k}^{\left[
j\right]  }\left(  \theta\right)  =\frac{\left(  2k-2\right)  !}{4^{k-1}}%
\sum_{m=2k}^{2j}\left(  \frac{2^{m-2}}{\left(  m-2\right)  !}\left\vert
t\left(  m,2k\right)  \right\vert \sin^{m-2}\left(  \theta/2\right)
-\frac{2^{m}}{m!}\frac{m^{2}}{4}\left\vert t\left(  m,2k\right)  \right\vert
\sin^{m}\left(  \theta/2\right)  \right)  \ .
\end{equation}
Then compare this to (\ref{IntegerSpinEvenCoef}) for $2k\rightarrow2k-2$,
after rewriting the latter by making use of the elementary recurrence formulas
for the central factorial numbers, as given in Proposition 2.1 of \cite{CFN},
say. \ In particular, $t\left(  m,n\right)  =0$ for $m<n$, and otherwise\
\begin{equation}
t\left(  m,2k-2\right)  =t\left(  m+2,2k\right)  +\frac{1}{4}~m^{2}~t\left(
m,2k\right)  \ .
\end{equation}
Using this recurrence relation in (\ref{IntegerSpinEvenCoef}) gives
\begin{align}
A_{2k-2}^{\left[  j\right]  }\left(  \theta\right)   &  =\frac{\left(
2k-2\right)  !}{\left(  2j\right)  !}\frac{4^{j}}{4^{k-1}}\left\vert t\left(
2+2j,2k\right)  \right\vert \sin^{2j}\left(  \theta/2\right)  +\frac{\left(
2k-2\right)  !}{4^{k-1}}\sum_{m=2k}^{2j}\frac{2^{m-2}}{\left(  m-2\right)
!}\left\vert t\left(  m,2k\right)  \right\vert \sin^{m-2}\left(
\theta/2\right) \nonumber\\
&  -\frac{\left(  2k-2\right)  !}{4^{k-1}}\sum_{m=2k}^{2j}\frac{2^{m}}%
{m!}\frac{m^{2}}{4}\left\vert t\left(  m,2k\right)  \right\vert \sin
^{m}\left(  \theta/2\right)  \ ,
\end{align}
upon assigning the correct phases (see (\ref{A2}) in Appendix A). \ But the
highest coefficient $A_{2j}^{\left[  j\right]  }$ is readily shown to be (e.g.
see \cite{CFZ})
\begin{equation}
A_{2j}^{\left[  j\right]  }\left(  \theta\right)  =\sin^{2j}\left(
\theta/2\right)  \ .
\end{equation}
Therefore (\ref{2ndOrderDiffEqn}) and (\ref{IntegerSpinEvenCoef}) are
verified. \ 

For semi-integer $j$ a similar derivation involving the odd index coefficients
$A_{2k+1}^{\left[  j\right]  }\left(  \theta\right)  $ goes through perfectly
in parallel to the even index case, and thereby completes this derivation of
the CFZ results.

\section*{A Derivation Using Biorthogonality}

Considerable analysis and combinatorics are implicit in (\ref{the result}) and
(\ref{the coefficients}). \ Perhaps the analytic features of the CFZ formulas
are most fully appreciated if viewed in the context of biorthogonal systems.
\ In any case, the theory of biorthogonal systems naturally leads to another
proof of (\ref{the result}).

\subsection*{Biorthogonal Functions}

Since the Taylor polynomials produced by the truncation in
(\ref{the coefficients}) involve only even powers of $\sin\left(
\theta/2\right)  $ with non-negative coefficients, the resulting set of
polynomials are \emph{not} orthogonal for any positive measure on $\theta$.
\ Instead, the dual function space consists of linear combinations of
Chebyshev polynomials ($\cos\left(  k\theta\right)  $ with $0\leq k\leq j$,
for any fixed integer value of $j$) and these linear combinations alternate in
sign as functions of $\theta$ to give the requisite orthogonality. \ This
provides, in a quantum physics context, an elementary example of a finitely
indexed \emph{biorthogonal system of functions} \cite{Young}. \ (For examples
of infinite and countably indexed biorthogonal quantum systems, see
\cite{CM,CMS}. \ For other examples and a careful discussion of the relevant
theory, see \cite{Brody}.)

To understand this structure, consider a basis of monomials of \emph{even}
powers of $\sin\left(  \theta/2\right)  $. \ The $j+1$ lowest powers of
$\sin^{2}\left(  \theta/2\right)  $, beginning with $1=\sin^{0}\left(
\theta/2\right)  $ and ending with $\sin^{2j}\left(  \theta/2\right)  $,
constitute one half of a\emph{\ finite} biorthogonal system of functions $f$
and their duals $g$: $\left\{  f_{n}^{\left[  j\right]  },~g_{n}^{\left[
j\right]  }~|~n=0,1,\cdots,j\right\}  $. \ For integer $j$ the functions and
their duals are given by
\begin{subequations}
\begin{align}
f_{0}^{\left[  j\right]  }\left(  \theta\right)   &  =1\text{ \ \ and
\ \ }g_{0}^{\left[  j\right]  }\left(  \theta\right)  =1+2\sum_{k=1}^{j}%
\cos\left(  k\theta\right)  \text{ \ \ for }n=0\ ,\label{DualFunctions0}\\
f_{n}^{\left[  j\right]  }\left(  \theta\right)   &  =\sin^{2n}\left(
\theta/2\right)  \text{ \ \ and \ \ }g_{n}^{\left[  j\right]  }\left(
\theta\right)  =\left(  -4\right)  ^{n}\sum_{k=n}^{j}\frac{k}{n}\binom
{k+n-1}{2n-1}\cos\left(  k\theta\right)  \text{ \ \ for }0<n\leq j\ .
\label{DualFunctions}%
\end{align}
These are orthogonal and normalized for any particular $j$:
\end{subequations}
\begin{equation}
\delta_{m,n}=\frac{1}{2\pi}\int_{-\pi}^{+\pi}g_{m}^{\left[  j\right]  }\left(
\theta\right)  f_{n}^{\left[  j\right]  }\left(  \theta\right)  d\theta\ .
\label{OrthoNorm}%
\end{equation}
In the spirit of (\ref{the coefficients}) the dual functions may be written as
truncations of infinite series in the variable $w=e^{i\theta}$, namely,%
\begin{equation}
g_{n}^{\left[  j\right]  }\left(  \theta\right)  =\operatorname{Re}\left(
h_{n}^{\left[  j\right]  }\left(  \theta\right)  \right)  \text{ \ \ where
\ \ }h_{n}^{\left[  j\right]  }\left(  \theta\right)  =\operatorname*{Trunc}%
_{j}\left[  \frac{\left(  -4w\right)  ^{n}\left(  1+w\right)  }{\left(
1-w\right)  ^{1+2n}}\right]  _{w=e^{i\theta}}\text{ \ \ for\ }0<n\leq j.
\label{DualTruncs}%
\end{equation}

There are similar results for a finite biorthogonal system of functions
consisting of the \emph{odd} powers $\sin^{2n-1}\left(  \theta/2\right)  $,
and their duals. \ This other system is easily obtained from the\ biorthogonal
system involving the even powers of $\sin\left(  \theta/2\right)  $, as given
above, just by moving a single factor of $\sin\left(  \theta/2\right)  $ from
the functions to the dual functions. \ For application to the spin matrix
expansion, consider semi-integer $j$. \ The $j+1/2$ lowest odd powers of
$\sin\left(  \theta/2\right)  $, beginning with $\sin\left(  \theta/2\right)
$ and ending with $\sin^{2j}\left(  \theta/2\right)  $, again constitute one
half of a biorthogonal system of functions $f$ and their duals $g$: $\left\{
f_{n}^{\left[  j\right]  },~g_{n}^{\left[  j\right]  }~|~n=1,\cdots
,j+1/2\right\}  $. \ For semi-integer $j$ the functions and their duals are
now given by
\begin{equation}
f_{n}^{\left[  j\right]  }\left(  \theta\right)  =\sin^{2n-1}\left(
\theta/2\right)  \text{ \ \ and \ \ }g_{n}^{\left[  j\right]  }\left(
\theta\right)  =\left(  -4\right)  ^{n}\sin\left(  \theta/2\right)  \sum
_{k=n}^{j+1/2}\frac{k}{n}\binom{k+n-1}{2n-1}\cos\left(  k\theta\right)  \text{
\ \ for }1\leq n\leq j+1/2\ . \label{OddSystem}%
\end{equation}
Once again, these are orthogonal and normalized as in (\ref{OrthoNorm}).
\ Note that the dual functions of the latter system are orthogonal to all
\emph{even} powers of $\sin\left(  \theta/2\right)  $. \ Equivalently, the
dual functions of the biorthogonal system discussed previously are orthogonal
to all odd powers of $\sin\left(  \theta/2\right)  $. \ Thus the two systems
may be combined into a larger one, involving both even and odd powers of
$\sin\left(  \theta/2\right)  $, without modification of the dual functions.
\ To be even more explicit, additional details are given in Appendix D,
including some useful Tables.

\subsection*{Biorthogonal Matrices}

Next, consider \emph{dual matrices} which are trace orthonormalized with
respect to powers of the spin matrix, $S\equiv2~\boldsymbol{\hat{n}\cdot J}$.
\ Without loss of generality, choose $S=2J_{3}$, since any other choice for
$\boldsymbol{\hat{n}}$\ merely requires selecting a different basis to
diagonalize the spin matrix, thereby obtaining the same eigenvalues as
$2J_{3}$. \ Thus the powers are
\begin{equation}
S^{m}=\left(
\begin{array}
[c]{ccccc}%
\left(  2j\right)  ^{m} & 0 & \cdots & 0 & 0\\
0 & \left(  2j-2\right)  ^{m} & \cdots & 0 & 0\\
\vdots & \vdots & \ddots & \vdots & \vdots\\
0 & 0 & \cdots & \left(  -2j+2\right)  ^{m} & 0\\
0 & 0 & \cdots & 0 & \left(  -2j\right)  ^{m}%
\end{array}
\right)  \ .
\end{equation}
Now construct orthonormalized dual matrices $T_{n}$ such that%
\begin{equation}
\delta_{n,m}=\operatorname*{Trace}\left(  T_{n}~S^{m}\right)
\ ,\ \ \ n,m=0,1,\cdots,2j\ . \label{OrthoTrace}%
\end{equation}
Clearly, the $T_{n}$\ may also be chosen to be diagonal $\left(  2j+1\right)
\times\left(  2j+1\right)  $ matrices in the basis that diagonalizes $S$. \ In
fact, for any spin $j$ the required entries on the diagonal of $T_{n}$ are
just the entries in the $\left(  n+1\right)  $st row of the inverted
Vandermonde matrix, $V^{-1}\left[  j\right]  $. \ (Note that here, unlike the
conventions in \cite{CFZ}, both rows and columns of the Vandermonde matrix and
its inverse are indexed as $1,2,\cdots,2j+1$.) \ That is, \
\begin{equation}
\left(  T_{n-1}\right)  _{kk}=\left(  V^{-1}\left[  j\right]  \right)
_{n,k}\ ,\ \ \ n,k=1,\cdots,2j+1\ . \label{DualMatrices}%
\end{equation}
This result follows immediately from the fact that the diagonal entries for
$S^{m}$ are just the entries in the corresponding column (i.e. the $\left(
m+1\right)  $st column) of the Vandermonde matrix, $V\left[  j\right]  $.%
\begin{equation}
V\left[  j\right]  =\left(
\begin{array}
[c]{ccccc}%
1 & 2j & \left(  2j\right)  ^{2} & \cdots & \left(  2j\right)  ^{2j}\\
1 & 2j-2 & \left(  2j-2\right)  ^{2} & \cdots & \left(  2j-2\right)  ^{2j}\\
\vdots & \vdots & \vdots & \ddots & \vdots\\
1 & -2j & \left(  -2j\right)  ^{2} & \cdots & \left(  -2j\right)  ^{2j}%
\end{array}
\right)  \ .
\end{equation}

Thus the effective metric, $G$, on the space spanned by powers of the spin $j$
matrices, defined such that
\begin{equation}
\delta_{m,n}=S^{m}\cdot G\cdot S^{n}\equiv\sum_{k,l}\left(  S^{m}\right)
_{kk}~G_{kl}~\left(  S^{n}\right)  _{ll}\ , \label{OrthoMetric}%
\end{equation}
is given by $G=\left(  V^{-1}\right)  ^{\dag}\left(  V^{-1}\right)  $. \ That
is to say, since $V$ and $V^{-1}$ are real in the chosen basis, the metric is
\begin{equation}
G_{kl}\left[  j\right]  =\sum_{i}V^{-1}\left[  j\right]  _{ik}V^{-1}\left[
j\right]  _{il}\ .
\end{equation}
Another way to write the orthonormality (\ref{OrthoMetric}) is by
incorporating the metric $G$ into a matrix trace.%
\begin{equation}
\delta_{m,n}=\operatorname*{Trace}\left(  B~S^{m}~G~S^{n}\right)  \ ,
\label{TraceNorm}%
\end{equation}
where $B$ is a singular matrix with \emph{all} entries equal to $1$. \ 

We have more to say about (\ref{TraceNorm}) in Appendix F, but first we
encourage the reader to consider the explicit examples of spin matrix powers,
their duals, the corresponding Vandermonde matrix and its inverse
$V^{-1}\left[  j\right]  $, and the metric $G$, for $j=1/2,\ 1,\ 3/2,$ and
$2$, as given in Appendix E.

\subsection*{Extracting the Coefficients}

Returning to the problem at hand, for any given spin $j$, the dual matrices
may be used to extract\ the individual angle-dependent coefficients in the
expansion of the rotation matrix, (\ref{the result}), in the basis that
diagonalizes $\boldsymbol{\hat{n}\cdot J}$. \ That is,%
\begin{equation}
A_{k}^{\left[  j\right]  }\left(  \theta\right)  =\left(  -i\right)
^{k}k!\operatorname*{Trace}\left[  T_{k}~e^{i\theta\left(  \boldsymbol{\hat
{n}}\cdot\boldsymbol{J}\right)  }\right]  =\left(  -i\right)  ^{k}%
k!\operatorname*{Trace}\left[  T_{k}~\left(
\begin{array}
[c]{cccc}%
e^{ij\theta} & 0 & \cdots & 0\\
0 & e^{i\left(  j-1\right)  \theta} & \cdots & 0\\
\vdots & \vdots & \ddots & \vdots\\
0 & 0 & \cdots & e^{-ij\theta}%
\end{array}
\right)  \right]  \ . \label{TraceProjection}%
\end{equation}
Consider for now only integer $j$ and \emph{even} $k\in\left\{  0,\cdots
,2j\right\}  $. \ From (\ref{TraceProjection}) the form
\begin{equation}
\left.  A_{k}^{\left[  j\right]  }\left(  \theta\right)  \right\vert
_{\substack{j\text{ integer}\\k\text{ even}}}=\sum_{n=k/2}^{j}a_{k,n}^{\left[
j\right]  }~\sin^{2n}\left(  \theta/2\right)  \label{bSeries}%
\end{equation}
may be argued to hold from generic behavior of the coefficients under Fourier
analysis (periodicity in $\theta$, symmetry under reflections, etc.). \ More
specifically, for integer $j$ and even $k\in\left\{  0,\cdots,2j\right\}  $,
with $n\in\left\{  k/2,\cdots,j\right\}  $, the previous dual functions for
even powers of $\sin\left(  \theta/2\right)  $ may be used to extract the
coefficients $a_{k,n}^{\left[  j\right]  }$ as%
\begin{align}
a_{k,n}^{\left[  j\right]  }  &  =\frac{1}{2\pi}\int_{-\pi}^{+\pi}%
g_{n}^{\left[  j\right]  }\left(  \theta\right)  A_{k}^{\left[  j\right]
}\left(  \theta\right)  d\theta\label{AngularProjection}\\
&  =\frac{\left(  -i\right)  ^{k}k!}{2\pi}\operatorname*{Trace}\left[
T_{k}~\left(
\begin{array}
[c]{cccc}%
\int_{-\pi}^{+\pi}g_{n}^{\left[  j\right]  }\left(  \theta\right)
e^{ij\theta}d\theta & 0 & \cdots & 0\\
0 & \int_{-\pi}^{+\pi}g_{n}^{\left[  j\right]  }\left(  \theta\right)
e^{i\left(  j-1\right)  \theta}d\theta & \cdots & 0\\
\vdots & \vdots & \ddots & \vdots\\
0 & 0 & \cdots & \int_{-\pi}^{+\pi}g_{n}^{\left[  j\right]  }\left(
\theta\right)  e^{-ij\theta}d\theta
\end{array}
\right)  \right]  \ .\nonumber
\end{align}
As established in (\ref{DualMatrices}), the diagonal elements $\left(
T_{k}\right)  _{n,n}$ are just the entries in the $n$th column of the $\left(
k+1\right)  $st row of the inverted Vandermonde matrix, $V^{-1}\left[
j\right]  $. \ So,%
\begin{equation}
a_{k,n}^{\left[  j\right]  }=\left(  -i\right)  ^{k}k!\sum_{m=-j}^{j}\left(
V^{-1}\left[  j\right]  \right)  _{k+1,j-m+1}\frac{1}{2\pi}\int_{-\pi}^{+\pi
}g_{n}^{\left[  j\right]  }\left(  \theta\right)  e^{im\theta}d\theta\ .
\end{equation}
Again note that here, unlike the conventions in \cite{CFZ}, both the rows and
the columns of the Vandermonde matrix and its inverse are indexed as
$1,2,\cdots,\left(  2j+1\right)  $. \ 

Making use of (\ref{DualFunctions0}) for $n=0$ gives%
\begin{equation}
a_{k,0}^{\left[  j\right]  }=\left(  -i\right)  ^{k}k!\sum_{m=-j}^{j}\left(
V^{-1}\left[  j\right]  \right)  _{k+1,j-m+1}=\delta_{k,0}\ . \label{n=0 case}%
\end{equation}
It is true in general that the entries for any given row of the inverted
Vandermonde matrix sum to zero, \emph{except} for the first row, whose entries
sum to one. \ For instance, see the examples of $V^{-1}$ for
$j=1/2,\ 1,\ 3/2,$ and $2$ as given in Appendix E. \ We leave to the
interested reader the proof of this elementary fact for any $j$. \ 

Moreover, making use of (\ref{DualFunctions}) for $0<n\in\left\{
k/2,\cdots,j\right\}  $ gives%
\begin{align}
b_{k,n}^{\left[  j\right]  }  &  =\left(  -4\right)  ^{n}\left(  -i\right)
^{k}k!\sum_{m=-j}^{j}\left(  V^{-1}\left[  j\right]  \right)  _{k+1,j-m+1}%
\sum_{\ell=n}^{j}\frac{\ell}{n}\binom{\ell+n-1}{2n-1}\frac{1}{2\pi}\int_{-\pi
}^{+\pi}\cos\left(  \ell\theta\right)  e^{im\theta}d\theta\nonumber\\
&  =\left(  -4\right)  ^{n}\left(  -i\right)  ^{k}k!\sum_{m=-j}^{j}\left(
V^{-1}\left[  j\right]  \right)  _{k+1,j-m+1}\sum_{\ell=n}^{j}\frac{\ell}%
{n}\binom{\ell+n-1}{2n-1}\frac{1}{2}\left(  \delta_{-\ell,m}+\delta_{\ell
,m}\right) \nonumber\\
&  =\frac{1}{2}\left(  -4\right)  ^{n}\left(  -i\right)  ^{k}k!\sum_{\ell
=n}^{j}\frac{\ell}{n}\binom{\ell+n-1}{2n-1}\sum_{m=-j}^{j}\left(
V^{-1}\left[  j\right]  \right)  _{k+1,j-m+1}\left(  \delta_{-\ell,m}%
+\delta_{\ell,m}\right) \nonumber
\end{align}%
\begin{align}
&  =\frac{1}{2}\left(  -4\right)  ^{n}\left(  -i\right)  ^{k}k!\sum_{\ell
=n}^{j}\frac{\ell}{n}\binom{\ell+n-1}{2n-1}\left(  \left(  V^{-1}\left[
j\right]  \right)  _{k+1,j+\ell+1}+\left(  V^{-1}\left[  j\right]  \right)
_{k+1,j-\ell+1}\right) \nonumber\\
&  =\frac{\left(  -4\right)  ^{n}\left(  -1\right)  ^{k/2}k!}{n}\sum_{\ell
=n}^{j}\ell\binom{\ell+n-1}{2n-1}\left(  V^{-1}\left[  j\right]  \right)
_{k+1,j-\ell+1}\ . \label{almost there}%
\end{align}
In this last relation, the fact that the odd rows of $V^{-1}$ are
left$\leftrightarrow$right column symmetric was used. \ For instance, again
see the examples of $V^{-1}$ for $j=1/2,\ 1,\ 3/2,$ and $2$ as given in
Appendix E. \ Also note the even rows of $V^{-1}$ are left$\leftrightarrow
$right column antisymmetric. \ Again we leave the proof of these elementary
facts to the interested reader.

So for even $k\leq2j$, upon shifting the summation variable $\ell=j+1-m$,
(\ref{almost there}) becomes%
\begin{equation}
a_{k,n}^{\left[  j\right]  }=\left(  -4\right)  ^{n}\left(  -1\right)
^{k/2}k!\sum_{m=1}^{j+1-n}\left(  V^{-1}\left[  j\right]  \right)
_{k+1,m}\frac{\left(  j+1-m\right)  }{n}\binom{j+n-m}{2n-1}\ .
\label{even better}%
\end{equation}
Another Lemma involving the central factorial numbers and the inverted
Vandermonde matrix now comes into play, namely,
\begin{equation}
t\left(  2n,2l\right)  =\left(  2n\right)  !~2^{2l}\sum_{m=1}^{j+1-n}\left(
V^{-1}\left[  j\right]  \right)  _{2l+1,m}\frac{\left(  j+1-m\right)  }%
{n}\binom{j+n-m}{2n-1}\ ,
\end{equation}
for $l\in\left\{  1,\cdots,j\right\}  $ and $n\in\left\{  l,\cdots,j\right\}
$. \ This is established in Appendix C. \ As a result of this Lemma,
(\ref{even better}), and the special case (\ref{n=0 case}), it follows that
for even $k\in\left\{  0,\cdots,2j\right\}  $,%
\begin{equation}
a_{k,n}^{\left[  j\right]  }=\frac{2^{2n}}{2^{k}}\frac{k!}{\left(  2n\right)
!}\left\vert t\left(  2n,k\right)  \right\vert \ \text{\ \ \ for \ \ }%
2n\in\left\{  k,\cdots,2j\right\}  \ ,
\end{equation}
where various phases have been cancelled (once more see (\ref{A2}) in Appendix
A). \ 

Note that \textit{the only dependence on }$j$ is in the upper limit of the sum
involved in the series (\ref{bSeries}), and, correspondingly, on the allowed
values of $k$ and $n$ for a given $j$ . \ There is no \emph{explicit} $j$
dependence in any of the $a_{k,n}^{\left[  j\right]  }$ coefficients. \ 

So for integer $j$ and even $k$,
\begin{equation}
A_{k}^{\left[  j\right]  }\left(  \theta\right)  =\sum_{n=k/2}^{j}\frac
{2^{2n}}{2^{k}}\frac{k!}{\left(  2n\right)  !}\left\vert t\left(  2n,k\right)
\right\vert \sin^{2n}\left(  \theta/2\right)  =\frac{k!}{2^{k}}\sum_{m=k}%
^{2j}\frac{2^{m}}{m!}\left\vert t\left(  m,k\right)  \right\vert \sin
^{m}\left(  \theta/2\right)  \ ,
\end{equation}
in agreement with (\ref{IntegerSpinEvenCoef}). \ 

The remaining terms for integer $j$, i.e. the odd $k$ cases in
(\ref{the coefficients}), again follow from the fact \cite{CFZ} that odd
$k=2m-1$ coefficients are obtained by differentiating even $k=2m$
coefficients, as in (\ref{oddbydiff}). \ Thus for integer $j$ this
biorthogonal-system-based derivation of (\ref{the result}) is complete. \ 

For semi-integer $j$, a parallel derivation can be constructed using the
biorthogonal system of functions involving the odd powers $\sin^{2n+1}\left(
\theta/2\right)  $ and their duals. \ The details are left as an exercise for
the reader.

\section*{Conclusion}

The results of Curtright, Fairlie, and Zachos --- for rotations expressed as
polynomials of spin matrices --- were derived here in careful detail, first by
elementary methods that rely on the differential relations obtained in
\cite{CFZ}, and then by methods from the theory of biorthogonal systems, where
properties of the central factorial numbers were invoked in both derivations.
\ Either approach confirms the elegant expressions (\ref{the result}) and
(\ref{the coefficients}).

\paragraph*{Acknowledgement}

We wish to thank D Fairlie and C Zachos for discussions related to this work.
\ We also thank Jack and Peggy Nichols for their encouragement and support,
and especially for their hospitality, while this paper was in preparation.
\ Finally, we thank an anonymous reviewer for pointing out a significant
typographical error, and for bringing to our attention references
\cite{vanW}\ and \cite{WW}. \ This work was supported in part by NSF Award
PHY-1214521, and in part by a University of Miami Cooper Fellowship.

\newpage

\subsection*{Appendix A: \ Central factorial numbers}

For historical reasons, central factorial numbers are defined as the
coefficients in simple polynomials \cite{Riordan,CFN}. \ They can be either
positive or negative, but only their absolute values are needed for the
coefficients of the spin matrix expansions in the main text. Moreover,
$t\left(  even,even\right)  $ are integers, but $t\left(  odd,odd\right)  $
are not integers, and $t\left(  odd,even\right)  =0=t\left(  even,odd\right)
$. \ So the even and odd cases are best handled separately. \ 

By definition and as elementary consequences thereof (cf.
\emph{http://oeis.org/A182867}),%
\begin{gather}
\prod\limits_{l=0}^{m-1}\left(  x^{2}-l^{2}\right)  =\sum_{k=1}^{m}t\left(
2m,2k\right)  x^{2k}\ ,\tag{A1}\label{A1}\\
t\left(  2m,2k\right)  =\left(  -1\right)  ^{m-k}\left\vert t\left(
2m,2k\right)  \right\vert =\left(  -1\right)  ^{m-k}\frac{1}{k!}\left.
\frac{d^{k}}{dz^{k}}\prod\limits_{l=0}^{m-1}\left(  z+l^{2}\right)
\right\vert _{z=0}\ , \tag{A2}\label{A2}%
\end{gather}
as well as (cf. \emph{http://oeis.org/A008956})%
\begin{gather}
x\prod\limits_{l=0}^{m-1}\left(  x^{2}-\left(  l+\frac{1}{2}\right)
^{2}\right)  =\sum_{k=0}^{m}t\left(  2m+1,2k+1\right)  x^{2k+1}\ ,
\tag{A4}\label{A4}\\
t\left(  2m+1,2k+1\right)  =\left(  -1\right)  ^{m-k}\left\vert t\left(
2m+1,2k+1\right)  \right\vert =\left(  -1\right)  ^{m-k}\frac{1}{k!}\left.
\frac{d^{k}}{dz^{k}}\prod\limits_{l=0}^{m-1}\left(  z+\left(  l+\frac{1}%
{2}\right)  ^{2}\right)  \right\vert _{z=0}\ . \tag{A5}%
\end{gather}

\subsection*{Appendix B: \ A useful lemma}

Proof of the Lemma:%
\begin{equation}
\left(  2\boldsymbol{\hat{n}\cdot J}\right)  ^{2j+1}=-\sum_{k=0}%
^{2j}2^{1+2j-k}\times t\left(  2+2j,1+k\right)  \times\left(
2\boldsymbol{\hat{n}\cdot J}\right)  ^{k}\ , \tag{B1}\label{LemmaResult}%
\end{equation}
where $t\left(  m,n\right)  $ are the central factorial numbers, defined in
Appendix A. \ In a basis where $2\boldsymbol{\hat{n}\cdot J}$ is diagonal,
(\ref{LemmaResult}) reduces to a matrix equation,
\begin{equation}
\left(
\begin{array}
[c]{c}%
\left(  2j\right)  ^{2j+1}\smallskip\\
\left(  2j-2\right)  ^{2j+1}\\
\vdots\smallskip\\
\left(  -2j+2\right)  ^{2j+1}\smallskip\\
\left(  -2j\right)  ^{2j+1}%
\end{array}
\right)  =-2^{1+2j}\times V\left[  j\right]  \left(
\begin{array}
[c]{c}%
t\left(  2+2j,1\right)  \smallskip\\
\frac{1}{2}~t\left(  2+2j,2\right) \\
\vdots\smallskip\\
\frac{1}{2^{2j-1}}~t\left(  2+2j,2j\right)  \smallskip\\
\frac{1}{2^{2j}}~t\left(  2+2j,1+2j\right)
\end{array}
\right)  \ , \tag{B2}\label{LemmaRHS}%
\end{equation}
where the Vandermonde matrix for spin $j$ is defined by%
\begin{equation}
V\left[  j\right]  =\left(
\begin{array}
[c]{ccccc}%
1 & 2j & \left(  2j\right)  ^{2} & \cdots & \left(  2j\right)  ^{2j}\\
1 & 2j-2 & \left(  2j-2\right)  ^{2} & \cdots & \left(  2j-2\right)  ^{2j}\\
\vdots & \vdots & \vdots & \ddots & \vdots\\
1 & -2j & \left(  -2j\right)  ^{2} & \cdots & \left(  -2j\right)  ^{2j}%
\end{array}
\right)  \ . \tag{B3}%
\end{equation}
So, consider the $k$th row on the RHS of (\ref{LemmaRHS}): \
\begin{equation}
-\frac{2^{1+2j}}{\left(  j+1-k\right)  }\times\left(
\begin{array}
[c]{c}%
t\left(  2+2j,1\right)  \left(  j+1-k\right)  +t\left(  2+2j,2\right)  \left(
j+1-k\right)  ^{2}+t\left(  2+2j,3\right)  \left(  j+1-k\right)  ^{3}+\cdots\\
+t\left(  2+2j,2j\right)  \left(  j+1-k\right)  ^{2j-1}+t\left(
2+2j,1+2j\right)  \left(  j+1-k\right)  ^{2j}%
\end{array}
\right)  \ . \tag{B4}%
\end{equation}

If $j$ is an integer, then $t\left(  2+2j,odd\right)  =0$, and this $k$th row
becomes%
\begin{align}
&  -\frac{2^{1+2j}}{\left(  j+1-k\right)  }\times\left(
\begin{array}
[c]{c}%
t\left(  2+2j,2\right)  \left(  j+1-k\right)  ^{2}+t\left(  2+2j,4\right)
\left(  j+1-k\right)  ^{4}+\cdots\\
+t\left(  2+2j,2j-2\right)  \left(  j+1-k\right)  ^{2j-2}+t\left(
2+2j,2j\right)  \left(  j+1-k\right)  ^{2j}\\
+t\left(  2+2j,2j+2\right)  \left(  j+1-k\right)  ^{2j+2}-t\left(
2+2j,2j+2\right)  \left(  j+1-k\right)  ^{2j+2}%
\end{array}
\right) \nonumber\\
&  =-\frac{2^{1+2j}}{\left(  j+1-k\right)  }\times\left(  \prod\limits_{l=0}%
^{j}\left(  \left(  j+1-k\right)  ^{2}-l^{2}\right)  -t\left(
2+2j,2j+2\right)  \times\left(  j+1-k\right)  ^{2j+2}\right)  \text{\ ,}
\tag{B5}%
\end{align}
where the $t\left(  2+2j,2j+2\right)  $ term was added and subtracted to
obtain the complete sum on the RHS of (\ref{A1}), for $m=j+1$ and $x=j+1-k$,
and then that sum was replaced with the product on the LHS of (\ref{A1}).
\ But the product evaluates to zero because one of the terms in the product
always vanishes for $k\geq1$, and therefore the $k$th row on the RHS of
(\ref{LemmaRHS}) is%
\begin{equation}
t\left(  2+2j,2+2j\right)  \times\left(  2j+2-2k\right)  ^{2j+1}=\left(
2j+2-2k\right)  ^{2j+1}\ , \tag{B6}%
\end{equation}
since\ $t\left(  n,n\right)  =1$, \ Thus we obtain the $k$th row on the LHS of
(\ref{LemmaRHS}), and the Lemma is proven for integer $j$.

If $j$ is semi-integer, a corresponding proof goes through just as easily,
upon using (\ref{A4}).

\subsection*{Appendix C: \ Another useful lemma}

Proof of the Lemma:%
\begin{equation}
t\left(  2n,2l\right)  =\left(  2n\right)  !~2^{2l}\sum_{m=1}^{j+1-n}\left(
V^{-1}\left[  j\right]  \right)  _{2l+1,m}\frac{\left(  j+1-m\right)  }%
{n}\binom{j+n-m}{2n-1}\ , \tag{C1}%
\end{equation}
for integer $j$, for $l\in\left\{  1,\cdots,j\right\}  $, and for
$0<n\in\left\{  l,\cdots,j\right\}  $, where $t\left(  m,k\right)  $ are the
central factorial numbers defined in Appendix A. \ Following steps similar to
those used in Appendix B leads immediately to the result:%
\begin{equation}
\sum_{q=1}^{2j+1}\left(  V\left[  j\right]  \right)  _{m,q}\frac{1}{2^{q}%
}~t\left(  2n,q-1\right)  =\sum_{l=1}^{j}\left(  V\left[  j\right]  \right)
_{m,2l+1}\frac{1}{2^{2l+1}}~t\left(  2n,2l\right)  =\frac{1}{2}\prod
\limits_{k=0}^{n-1}\left(  \left(  j+1-m\right)  ^{2}-k^{2}\right)  \ .
\tag{C2}%
\end{equation}
It then follows from left-multiplication by $V^{-1}\left[  j\right]  $ that
\begin{equation}
\frac{1}{2^{2l+1}}~t\left(  2n,2l\right)  =\frac{1}{2}\sum_{m=1}^{2j+1}\left(
V^{-1}\left[  j\right]  \right)  _{2l+1,m}\prod\limits_{k=0}^{n-1}\left(
\left(  j+1-m\right)  ^{2}-k^{2}\right)  \ . \tag{C3}%
\end{equation}
These are the non-vanishing cases of interest. \ But then,
\begin{equation}
\prod\limits_{k=0}^{n-1}\left(  \left(  j+1-m\right)  ^{2}-k^{2}\right)
=\left\{
\begin{array}
[c]{ccc}%
\dfrac{\left(  j+1-m\right)  \left(  j+n-m\right)  !}{\left(  j+1-n-m\right)
!} & \text{for} & 1\leq m\leq j+1-n\\
0 & \text{for} & j+2-n\leq m\leq j+n\\
\dfrac{\left(  m-1-j\right)  \left(  m+n-j-2\right)  !}{\left(
m-n-j-1\right)  !} & \text{for} & j+1+n\leq m\leq2j+1
\end{array}
\right.  \ . \tag{C4}%
\end{equation}
Thus we obtain%
\begin{align}
t\left(  2n,2l\right)   &  =2^{2l}\sum_{m=1}^{j+1-n}\left(  V^{-1}\left[
j\right]  \right)  _{2l+1,m}\frac{\left(  j+1-m\right)  ~\left(  j+n-m\right)
!}{\left(  j+1-n-m\right)  !}\nonumber\\
&  +2^{2l}\sum_{m=j+1+n}^{2j+1}\left(  V^{-1}\left[  j\right]  \right)
_{2l+1,m}\frac{\left(  m-1-j\right)  \left(  m+n-j-2\right)  !}{\left(
m-n-j-1\right)  !}\ . \tag{C5}\label{TwoSums}%
\end{align}
However, the two sums on the RHS of (\ref{TwoSums}) are equal. \ In fact, the
summands are equal term-by-term, as a consequence of the left$\leftrightarrow
$right column symmetry of the odd rows of $V^{-1}\left[  j\right]  $. \ That
is to say,%
\begin{align}
\left.  \frac{\left(  j+1-k\right)  ~\left(  j+n-k\right)  !}{\left(
j+1-n-k\right)  !}\right\vert _{k=m}  &  =\left.  \frac{\left(  k-1-j\right)
\left(  k+n-j-2\right)  !}{\left(  k-n-j-1\right)  !}\right\vert
_{k=2j+2-m}\ ,\nonumber\\
& \nonumber\\
\left(  V^{-1}\left[  j\right]  \right)  _{2l+1,m}  &  =\left(  V^{-1}\left[
j\right]  \right)  _{2l+1,2j+2-m}\text{\ ,} \tag{C6}%
\end{align}
for $m=1,\cdots,j+1$. \ The final result is then the sought-for relation:%
\begin{align}
t\left(  2n,2l\right)   &  =2\times2^{2l}\sum_{m=1}^{j+1-n}\left(
V^{-1}\left[  j\right]  \right)  _{2l+1,m}\frac{\left(  j+1-m\right)  ~\left(
j+n-m\right)  !}{\left(  j+1-n-m\right)  !}\nonumber\\
&  \equiv\left(  2n\right)  !~2^{2l}\sum_{m=1}^{j+1-n}\left(  V^{-1}\left[
j\right]  \right)  _{2l+1,m}\frac{\left(  j+1-m\right)  }{n}\binom
{j+n-m}{2n-1}\ . \tag{C7}%
\end{align}
\newpage

\subsection*{Appendix D: \ Finite biorthogonal systems of functions}

Here are more details about the finite biorthogonal systems of functions
described in the main text, constructed such that \
\begin{equation}
\delta_{m,n}=\frac{1}{2\pi}\int_{-\pi}^{+\pi}g_{m}^{\left[  j\right]  }\left(
\theta\right)  f_{n}^{\left[  j\right]  }\left(  \theta\right)  d\theta\ .
\tag{D1}%
\end{equation}
Consider first the even powers of $\sin\left(  \theta/2\right)  $ and their
duals. \ The biorthogonal system in this situation is given by the following
Table.%
\begin{align*}
\text{{}}  &  \text{\hspace*{-0.3in}\textbf{Table 1 \ \ }For }j\in\left\{
0,1,2,3,\cdots\right\}  \text{:}\\
&  \hspace*{-0.3in}\fbox{$%
\begin{array}
[c]{ccc}%
\text{Function }f_{n}^{\left[  j\right]  } & \ \ \ \ \ n\ \ \ \ \ \vspace
{-0.1in} & \text{Dual Function }g_{n}^{\left[  j\right]  }\\
\hrulefill & \hrulefill\smallskip & \hrulefill\\
1 & 0\smallskip & \left[  1+2\cos\left(  \theta\right)  +2\cos\left(
2\theta\right)  +2\cos\left(  3\theta\right)  +2\cos\left(  4\theta\right)
+2\cos\left(  5\theta\right)  +\cdots+2\cos\left(  j\theta\right)  \right] \\
\sin^{2}\left(  \theta/2\right)  & 1\smallskip & -4\left[  \cos\left(
\theta\right)  +4\cos\left(  2\theta\right)  +9\cos\left(  3\theta\right)
+16\cos\left(  4\theta\right)  +25\cos\left(  5\theta\right)  +\cdots
+j^{2}\cos\left(  j\theta\right)  \right] \\
\sin^{4}\left(  \theta/2\right)  & 2\smallskip & 16\left[  \cos\left(
2\theta\right)  +6\cos\left(  3\theta\right)  +20\cos\left(  4\theta\right)
+50\cos\left(  5\theta\right)  +\cdots+\frac{1}{2}\binom{j+1}{3}j\cos\left(
j\theta\right)  \right] \\
\sin^{6}\left(  \theta/2\right)  & 3\smallskip & -64\left[  \cos\left(
3\theta\right)  +8\cos\left(  4\theta\right)  +35\cos\left(  5\theta\right)
+112\cos\left(  6\theta\right)  +\cdots+\frac{1}{3}\binom{j+2}{5}j\cos\left(
j\theta\right)  \right] \\
\sin^{8}\left(  \theta/2\right)  & 4\smallskip & 256\left[  \cos\left(
4\theta\right)  +10\cos\left(  5\theta\right)  +54\cos\left(  6\theta\right)
+210\cos\left(  7\theta\right)  +\cdots+\frac{1}{4}\binom{j+3}{7}j\cos\left(
j\theta\right)  \right] \\
\sin^{10}\left(  \theta/2\right)  & 5\smallskip & -1024\left[  \cos\left(
5\theta\right)  +12\cos\left(  6\theta\right)  +77\cos\left(  7\theta\right)
+352\cos\left(  8\theta\right)  +\cdots+\frac{1}{5}\binom{j+4}{9}j\cos\left(
j\theta\right)  \right] \\
\ \vdots & \vdots\smallskip & \vdots
\ \ \ \ \ \ \ \ \ \ \ \ \ \ \ \ \ \ \ \ \ \ \ \ \ \ \ \ \ \ \vdots
\ \ \ \ \ \ \ \ \ \ \ \ \ \ \ \ \ \ \ \ \ \ \ \ \ \ \ \ \ \ \vdots\\
\sin^{2j-6}\left(  \theta/2\right)  & j-3\smallskip & \left(  -4\right)
^{j-3}\left[  \cos\left(  \left(  j-3\right)  \theta\right)  +2\left(
j-2\right)  \cos\left(  \left(  j-2\right)  \theta\right)  +\cdots+\frac
{1}{j-3}\binom{2j-4}{2j-7}j\cos\left(  j\theta\right)  \right] \\
\sin^{2j-4}\left(  \theta/2\right)  & j-2\smallskip & \left(  -4\right)
^{j-2}\left[  \cos\left(  \left(  j-2\right)  \theta\right)  +2\left(
j-1\right)  \cos\left(  \left(  j-1\right)  \theta\right)  +\left(
2j-3\right)  j\cos\left(  j\theta\right)  \right] \\
\sin^{2j-2}\left(  \theta/2\right)  & j-1\smallskip & \left(  -4\right)
^{j-1}\left[  \cos\left(  \left(  j-1\right)  \theta\right)  +2j\cos\left(
j\theta\right)  \right] \\
\sin^{2j}\left(  \theta/2\right)  & j & \left(  -4\right)  ^{j}\left[
\cos\left(  j\theta\right)  \right]
\end{array}
$}%
\end{align*}
Consider next the odd powers of $\sin\left(  \theta/2\right)  $ and their
duals. \ As noted in the main text, a biorthogonal system for this situation
may be obtained from the system involving even powers of $\sin\left(
\theta/2\right)  $ just by moving a single factor of $\sin\left(
\theta/2\right)  $ from the functions to the dual functions. \ For application
to the spin matrix expansion, consider semi-integer $j$. \ The $j+1/2$ lowest
odd powers of $\sin\left(  \theta/2\right)  $, beginning with $\sin\left(
\theta/2\right)  $ and ending with $\sin^{2j}\left(  \theta/2\right)  $, again
constitute one half of a biorthogonal system of functions. \ The functions and
their duals are given by the following Table.%
\begin{align*}
\text{{}}  &  \text{\hspace*{-0.75in}\textbf{Table 2 \ \ }For \ \ }%
j\in\left\{  \tfrac{1}{2},\tfrac{3}{2},\tfrac{5}{2},\tfrac{7}{2}%
,\cdots\right\}  \text{:}\\
&  \hspace*{-0.75in}\fbox{$%
\begin{array}
[c]{ccc}%
\text{Function }f_{n}^{\left[  j\right]  } & \ \ \ \ \ n\ \ \ \ \ \vspace
{-0.1in} & \text{Dual Function }g_{n}^{\left[  j\right]  }\\
\hrulefill & \hrulefill\smallskip & \hrulefill\\
\sin\left(  \theta/2\right)  & 1\smallskip & -4\sin\left(  \theta/2\right)
\left[  \cos\left(  \theta\right)  +4\cos\left(  2\theta\right)  +9\cos\left(
3\theta\right)  +16\cos\left(  4\theta\right)  +25\cos\left(  5\theta\right)
+\cdots+\left(  j+\frac{1}{2}\right)  ^{2}\cos\left(  \left(  j+\frac{1}%
{2}\right)  \theta\right)  \right] \\
\sin^{3}\left(  \theta/2\right)  & 2\smallskip & 16\sin\left(  \theta
/2\right)  \left[  \cos\left(  2\theta\right)  +6\cos\left(  3\theta\right)
+20\cos\left(  4\theta\right)  +50\cos\left(  5\theta\right)  +\cdots+\frac
{1}{2}\binom{j+3/2}{3}\left(  j+\frac{1}{2}\right)  \cos\left(  \left(
j+\frac{1}{2}\right)  \theta\right)  \right] \\
\sin^{5}\left(  \theta/2\right)  & 3\smallskip & -64\sin\left(  \theta
/2\right)  \left[  \cos\left(  3\theta\right)  +8\cos\left(  4\theta\right)
+35\cos\left(  5\theta\right)  +112\cos\left(  6\theta\right)  +\cdots
+\frac{1}{3}\binom{j+5/2}{5}\left(  j+\frac{1}{2}\right)  \cos\left(  \left(
j+\frac{1}{2}\right)  \theta\right)  \right] \\
\sin^{7}\left(  \theta/2\right)  & 4\smallskip & 256\sin\left(  \theta
/2\right)  \left[  \cos\left(  4\theta\right)  +10\cos\left(  5\theta\right)
+54\cos\left(  6\theta\right)  +210\cos\left(  7\theta\right)  +\cdots
+\frac{1}{4}\binom{j+7/2}{7}\left(  j+\frac{1}{2}\right)  \cos\left(  \left(
j+\frac{1}{2}\right)  \theta\right)  \right] \\
\sin^{9}\left(  \theta/2\right)  & 5\smallskip & -1024\sin\left(
\theta/2\right)  \left[  \cos\left(  5\theta\right)  +12\cos\left(
6\theta\right)  +77\cos\left(  7\theta\right)  +352\cos\left(  8\theta\right)
+\cdots+\frac{1}{5}\binom{j+9/2}{9}\left(  j+\frac{1}{2}\right)  \cos\left(
\left(  j+\frac{1}{2}\right)  \theta\right)  \right] \\
\sin^{11}\left(  \theta/2\right)  & 6 & 4096\sin\left(  \theta/2\right)
\left[  \cos\left(  6\theta\right)  +14\cos\left(  7\theta\right)
+104\cos\left(  8\theta\right)  +546\cos\left(  9\theta\right)  +\cdots
+\frac{1}{6}\binom{j+11/2}{11}\left(  j+\frac{1}{2}\right)  \cos\left(
\left(  j+\frac{1}{2}\right)  \theta\right)  \right] \\
\ \vdots & \vdots\smallskip & \vdots
\ \ \ \ \ \ \ \ \ \ \ \ \ \ \ \ \ \ \ \ \ \ \ \ \ \ \ \ \ \ \vdots
\ \ \ \ \ \ \ \ \ \ \ \ \ \ \ \ \ \ \ \ \ \ \ \ \ \ \ \ \ \ \vdots\\
\sin^{2j-6}\left(  \theta/2\right)  & j-\frac{5}{2}\smallskip & \left(
-4\right)  ^{j-\frac{5}{2}}\sin\left(  \theta/2\right)  \left[  \cos\left(
\left(  j-\frac{5}{2}\right)  \theta\right)  +2\left(  j-\frac{3}{2}\right)
\cos\left(  \left(  j-\frac{3}{2}\right)  \theta\right)  +\cdots+\frac
{1}{j-\frac{5}{2}}\binom{2j-3}{2j-6}\left(  j+\frac{1}{2}\right)  \cos\left(
\left(  j+\frac{1}{2}\right)  \theta\right)  \right] \\
\sin^{2j-4}\left(  \theta/2\right)  & j-\frac{3}{2}\smallskip & \left(
-4\right)  ^{j-\frac{3}{2}}\sin\left(  \theta/2\right)  \left[  \cos\left(
\left(  j-\frac{3}{2}\right)  \theta\right)  +2\left(  j-\frac{1}{2}\right)
\cos\left(  \left(  j-\frac{1}{2}\right)  \theta\right)  +\left(  2j-2\right)
\left(  j+\frac{1}{2}\right)  \cos\left(  \left(  j+\frac{1}{2}\right)
\theta\right)  \right] \\
\sin^{2j-2}\left(  \theta/2\right)  & j-\frac{1}{2}\smallskip & \left(
-4\right)  ^{j-\frac{1}{2}}\sin\left(  \theta/2\right)  \left[  \cos\left(
\left(  j-\frac{1}{2}\right)  \theta\right)  +2\left(  j+\frac{1}{2}\right)
\cos\left(  \left(  j+\frac{1}{2}\right)  \theta\right)  \right] \\
\sin^{2j}\left(  \theta/2\right)  & j+\frac{1}{2} & \left(  -4\right)
^{j+\frac{1}{2}}\sin\left(  \theta/2\right)  \left[  \cos\left(  \left(
j+\frac{1}{2}\right)  \theta\right)  \right]
\end{array}
$}%
\end{align*}
As stated in the main text, these two systems may be combined into a larger
one, involving both even and odd powers of $\sin\left(  \theta/2\right)  $,
without modification of the dual functions. \ An explicit Table for the
enlarged system may be obtained just by interlacing the rows of Tables 1 and 2.

The procedure to obtain the dual functions is straightforward. \ For example,
start at the highest power of $\sin\left(  \theta/2\right)  $, namely,
$\sin^{2j}\left(  \theta/2\right)  $, as given in the last row of Table 1, for
which an obvious dual function is the Chebyshev polynomial $\cos\left(
j\theta\right)  $ with coefficient $\left(  -4\right)  ^{j}$, as is easily
verified. \ Note that $\cos\left(  j\theta\right)  $ is manifestly orthogonal
to all powers $\sin^{2n}\left(  \theta/2\right)  $ with $n<j$. \ Now consider
the next to highest power, namely, $\sin^{2j-2}\left(  \theta/2\right)  $, as
given in the next-to-last row of Table 1. \ The first term of its dual
function is just what it would be if one were considering the biorthogonal
system with $j$ reduced by $1$, namely, $\cos\left(  \left(  j-1\right)
\theta\right)  $ with coefficient $\left(  -4\right)  ^{j-1}$. \ But this term
alone is not orthogonal to $\sin^{2j}\left(  \theta/2\right)  $, so one must
add the higher harmonic $\cos\left(  j\theta\right)  $ with coefficient to
achieve the desired orthogonality. \ The higher harmonic is clearly orthogonal
to $\sin^{2j-2}\left(  \theta/2\right)  $ and all lower powers of $\sin
^{2}\left(  \theta/2\right)  $, so it does not contribute to the
orthonormalization integral for $\sin^{2j-2}\left(  \theta/2\right)  $. \ And
so it goes. \ The results for the duals of the lower powers of $\sin
^{2}\left(  \theta/2\right)  $ are iterated series of terms where, in any
particular row of Table 1, all but the highest harmonic are given by the same
terms as appear in the subsequent row of the Table (after replacing $j$ in the
earlier row by $j+1$ in the later row), and where the coefficient of the
highest harmonic in the earlier row, namely, $\cos\left(  j\theta\right)  $,
is determined by requiring orthogonality to $\sin^{2j}\left(  \theta/2\right)
$.

To carry out the construction of the dual functions, the following integral is
useful:
\begin{equation}
\int_{-\pi}^{\pi}\cos\left(  m\theta\right)  \sin^{2j}\left(  \theta/2\right)
d\theta=\frac{2\pi\left(  -1\right)  ^{m}}{4^{j}}\binom{2j}{j+m}\text{ \ \ for
\ \ }0\leq m\leq j\text{\ .} \tag{D2}%
\end{equation}
Note also the following generating functions for the coefficients inside the
square brackets in the various rows of Table 1:%
\begin{align}
\frac{1+x}{1-x}  &  =1+2x+2x^{2}+2x^{3}+2x^{4}+2x^{5}+2x^{6}+2x^{7}+O\left(
x^{8}\right)  \ ,\tag{D3}\\
\frac{1+x}{\left(  1-x\right)  ^{3}}  &  =1+4x+9x^{2}+16x^{3}+25x^{4}%
+36x^{5}+49x^{6}+64x^{7}+O\left(  x^{8}\right)  \ ,\nonumber\\
\frac{1+x}{\left(  1-x\right)  ^{5}}  &  =1+6x+20x^{2}+50x^{3}+105x^{4}%
+196x^{5}+336x^{6}+540x^{7}+O\left(  x^{8}\right)  \ ,\nonumber\\
\frac{1+x}{\left(  1-x\right)  ^{7}}  &  =1+8x+35x^{2}+112x^{3}+294x^{4}%
+672x^{5}+1386x^{6}+2640x^{7}+O\left(  x^{8}\right)  \ ,\nonumber\\
\frac{1+x}{\left(  1-x\right)  ^{9}}  &  =1+10x+54x^{2}+210x^{3}%
+660x^{4}+1782x^{5}+4290x^{6}+9438x^{7}+O\left(  x^{8}\right)  \ ,\nonumber\\
\frac{1+x}{\left(  1-x\right)  ^{11}}  &  =1+12x+77x^{2}+352x^{3}%
+1287x^{4}+4004x^{5}+11\,011x^{6}+27\,456x^{7}+O\left(  x^{8}\right)
\ ,\nonumber\\
\frac{1+x}{\left(  1-x\right)  ^{13}}  &  =1+14x+104x^{2}+546x^{3}%
+2275x^{4}+8008x^{5}+24\,752x^{6}+68\,952x^{7}+O\left(  x^{8}\right)
\ ,\nonumber
\end{align}%
\begin{align}
\frac{1+x}{\left(  1-x\right)  ^{2j-5}}  &  =1+2\left(  j-2\right)  x+\left(
2j-5\right)  \left(  j-1\right)  x^{2}+\frac{2}{3}\left(  2j-5\right)  \left(
j-2\right)  j~x^{3}+O\left(  x^{4}\right)  \ ,\tag{D4}\\
\frac{1+x}{\left(  1-x\right)  ^{2j-3}}  &  =1+2\left(  j-1\right)  x+\left(
2j-3\right)  j~x^{2}+O\left(  x^{3}\right)  \ ,\nonumber\\
\frac{1+x}{\left(  1-x\right)  ^{2j-1}}  &  =1+2j~x+O\left(  x^{2}\right)
\ ,\nonumber\\
\frac{1+x}{\left(  1-x\right)  ^{2j+1}}  &  =1+O\left(  x\right)  \ .\nonumber
\end{align}
These generating functions are a direct consequence of (\ref{DualTruncs}).
\ Finally, note of course that $\frac{1}{2}\binom{j+1}{3}j=\frac{1}{12}\left(
j^{2}-1\right)  j^{2}$, and $\frac{1}{j-3}\binom{2j-4}{2j-7}j=\frac{2}%
{3}\left(  2j-5\right)  \left(  j-2\right)  j$, etc., but in the Tables these
are expressed as binomial coefficients to display more clearly the general
pattern.\newpage

\subsection*{Appendix E: \ Biorthogonal matrix examples}

Here are more details about the biorthogonal systems of spin matrices
described in the main text, for $j=1/2,\ 1,\ 3/2,$ and $2$.

Spin $j=1/2$ is deceptively simple. \ The independent powers of the spin
matrix are%
\[
S^{0}=\left(
\begin{array}
[c]{cc}%
1 & 0\\
0 & 1
\end{array}
\right)  \ ,\ \ \ S^{1}=\left(
\begin{array}
[c]{cc}%
1 & 0\\
0 & -1
\end{array}
\right)  \ ,
\]
and the corresponding trace-orthonormal dual matrices are the same up to a
normalization factor. (NB This is \emph{not} true for any other $j$.)%
\[
T_{0}=\frac{1}{2}\left(
\begin{array}
[c]{cc}%
1 & 0\\
0 & 1
\end{array}
\right)  \ ,\ \ \ T_{1}=\frac{1}{2}\left(
\begin{array}
[c]{cc}%
1 & 0\\
0 & -1
\end{array}
\right)  \ ,\ \ \ \text{i.e. \ \ }V^{-1}=\frac{1}{2}\left(
\begin{array}
[c]{cc}%
1 & 1\\
1 & -1
\end{array}
\right)  \ ,\ \ \ G=\frac{1}{2}\left(
\begin{array}
[c]{cc}%
1 & 0\\
0 & 1
\end{array}
\right)  \ .
\]
Spin $j=1$ is a more interesting example. \ The independent powers of the spin
matrix are given by%
\[
S^{0}=\left(
\begin{array}
[c]{ccc}%
1 & 0 & 0\\
0 & 1 & 0\\
0 & 0 & 1
\end{array}
\right)  \ ,\ \ \ S^{1}=\left(
\begin{array}
[c]{ccc}%
2 & 0 & 0\\
0 & 0 & 0\\
0 & 0 & -2
\end{array}
\right)  \ ,\ \ \ S^{2}=\left(
\begin{array}
[c]{ccc}%
4 & 0 & 0\\
0 & 0 & 0\\
0 & 0 & 4
\end{array}
\right)  \ ,
\]
and the corresponding trace-orthonormal dual matrices, as well as $V^{-1}$ and
$G$, are given by
\begin{align*}
T_{0}  &  =\left(
\begin{array}
[c]{ccc}%
0 & 0 & 0\\
0 & 1 & 0\\
0 & 0 & 0
\end{array}
\right)  \ ,\ \ \ T_{1}=\frac{1}{4}\left(
\begin{array}
[c]{ccc}%
1 & 0 & 0\\
0 & 0 & 0\\
0 & 0 & -1
\end{array}
\right)  \ ,\ \ T_{3}=\frac{1}{8}\left(
\begin{array}
[c]{ccc}%
1 & 0 & 0\\
0 & -2 & 0\\
0 & 0 & 1
\end{array}
\right)  \ ,\\
& \\
V^{-1}  &  =\frac{1}{8}\left(
\begin{array}
[c]{ccc}%
0 & 8 & 0\\
2 & 0 & -2\\
1 & -2 & 1
\end{array}
\right)  \ ,\ \ \ G=\frac{1}{64}\left(
\begin{array}
[c]{ccc}%
5 & -2 & -3\\
-2 & 68 & -2\\
-3 & -2 & 5
\end{array}
\right)  \ .
\end{align*}
Spin $j=3/2$ is also interesting. \ The independent powers of the spin matrix
are given by%
\[
S^{0}=\left(
\begin{array}
[c]{cccc}%
1 & 0 & 0 & 0\\
0 & 1 & 0 & 0\\
0 & 0 & 1 & 0\\
0 & 0 & 0 & 1
\end{array}
\right)  \ ,\ \ \ S^{1}=\left(
\begin{array}
[c]{cccc}%
3 & 0 & 0 & 0\\
0 & 1 & 0 & 0\\
0 & 0 & -1 & 0\\
0 & 0 & 0 & -3
\end{array}
\right)  \ ,\ \ \ S^{2}=\left(
\begin{array}
[c]{cccc}%
9 & 0 & 0 & 0\\
0 & 1 & 0 & 0\\
0 & 0 & 1 & 0\\
0 & 0 & 0 & 9
\end{array}
\right)  \ ,\ \ \ S^{3}=\left(
\begin{array}
[c]{cccc}%
27 & 0 & 0 & 0\\
0 & 1 & 0 & 0\\
0 & 0 & -1 & 0\\
0 & 0 & 0 & -27
\end{array}
\right)  \ ,
\]
and the corresponding trace-orthonormal dual matrices, as well as $V^{-1}$ and
$G$, are given by%
\begin{gather*}
T_{0}=\frac{1}{16}\left(
\begin{array}
[c]{cccc}%
-1 & 0 & 0 & 0\\
0 & 9 & 0 & 0\\
0 & 0 & 9 & 0\\
0 & 0 & 0 & -1
\end{array}
\right)  \ ,\ \ \ T_{1}=\frac{1}{48}\left(
\begin{array}
[c]{cccc}%
-1 & 0 & 0 & 0\\
0 & 27 & 0 & 0\\
0 & 0 & -27 & 0\\
0 & 0 & 0 & 1
\end{array}
\right)  \ ,\ \ \ T_{2}=\frac{1}{16}\left(
\begin{array}
[c]{cccc}%
1 & 0 & 0 & 0\\
0 & -1 & 0 & 0\\
0 & 0 & -1 & 0\\
0 & 0 & 0 & 1
\end{array}
\right)  \ ,\\
\\
T_{3}=\frac{1}{48}\left(
\begin{array}
[c]{cccc}%
1 & 0 & 0 & 0\\
0 & -3 & 0 & 0\\
0 & 0 & 3 & 0\\
0 & 0 & 0 & -1
\end{array}
\right)  \ ,\ \ \ V^{-1}=\frac{1}{48}\left(
\begin{array}
[c]{cccc}%
-3 & 27 & 27 & -3\\
-1 & 27 & -27 & 1\\
3 & -3 & -3 & 3\\
1 & -3 & 3 & -1
\end{array}
\right)  \ .\ \ \ G=\frac{1}{48^{2}}\left(
\begin{array}
[c]{cccc}%
20 & -120 & -60 & 16\\
-120 & 1476 & 0 & -60\\
-60 & 0 & 1476 & -120\\
16 & -60 & -120 & 20
\end{array}
\right)  \ .
\end{gather*}
Spin $j=2$ helps to establish the general pattern.%
\[
S^{m}=\left(
\begin{array}
[c]{ccccc}%
4^{m} & 0 & 0 & 0 & 0\\
0 & 2^{m} & 0 & 0 & 0\\
0 & 0 & 0 & 0 & 0\\
0 & 0 & 0 & \left(  -2\right)  ^{m} & 0\\
0 & 0 & 0 & 0 & \left(  -4\right)  ^{m}%
\end{array}
\right)  \ ,\ \ \ T_{0}=\left(
\begin{array}
[c]{ccccc}%
0 & 0 & 0 & 0 & 0\\
0 & 0 & 0 & 0 & 0\\
0 & 0 & 1 & 0 & 0\\
0 & 0 & 0 & 0 & 0\\
0 & 0 & 0 & 0 & 0
\end{array}
\right)  \ ,\ \ \ T_{1}=\frac{1}{24}\left(
\begin{array}
[c]{ccccc}%
-1 & 0 & 0 & 0 & 0\\
0 & 8 & 0 & 0 & 0\\
0 & 0 & 0 & 0 & 0\\
0 & 0 & 0 & -8 & 0\\
0 & 0 & 0 & 0 & 1
\end{array}
\right)  \ ,\ \ \
\]%
\begin{gather*}
T_{2}=\frac{1}{96}\left(
\begin{array}
[c]{ccccc}%
-1 & 0 & 0 & 0 & 0\\
0 & 16 & 0 & 0 & 0\\
0 & 0 & -30 & 0 & 0\\
0 & 0 & 0 & 16 & 0\\
0 & 0 & 0 & 0 & -1
\end{array}
\right)  \ ,\ \ \ T_{3}=\frac{1}{96}\left(
\begin{array}
[c]{ccccc}%
1 & 0 & 0 & 0 & 0\\
0 & -2 & 0 & 0 & 0\\
0 & 0 & 0 & 0 & 0\\
0 & 0 & 0 & 2 & 0\\
0 & 0 & 0 & 0 & -1
\end{array}
\right)  \ ,\ \ \ T_{4}=\frac{1}{384}\left(
\begin{array}
[c]{ccccc}%
1 & 0 & 0 & 0 & 0\\
0 & -4 & 0 & 0 & 0\\
0 & 0 & 6 & 0 & 0\\
0 & 0 & 0 & -4 & 0\\
0 & 0 & 0 & 0 & 1
\end{array}
\right)  \ ,\\
\\
V^{-1}=\frac{1}{384}\left(
\begin{array}
[c]{ccccc}%
0 & 0 & 384 & 0 & 0\\
-16 & 128 & 0 & -128 & 16\\
-4 & 64 & -120 & 64 & -4\\
4 & -8 & 0 & 8 & -4\\
1 & -4 & 6 & -4 & 1
\end{array}
\right)  \ ,\ \ \ G=\frac{1}{384^{2}}\left(
\begin{array}
[c]{ccccc}%
289 & -2340 & 486 & 1820 & -255\\
-2340 & 20\,560 & -7704 & -12\,336 & 1820\\
486 & -7704 & 161\,892 & -7704 & 486\\
1820 & -12\,336 & -7704 & 20\,560 & -2340\\
-255 & 1820 & 486 & -2340 & 289
\end{array}
\right)  \ .
\end{gather*}

For the discussion in the text about the $S$ and $T$ matrix biorthogonality
expressed as a trace, and the related discussion in the following Appendix F,
it may also be helpful for the reader to consider the $B$ and $P$ matrices for
these spins, especially to check (\ref{BAsConjugatedProjector}) and
(\ref{PAsConjugatedB}). \ 

For example, for spin $j=1$,
\begin{align*}
B  &  =\left(
\begin{array}
[c]{c}%
1\\
1\\
1
\end{array}
\right)  \left(
\begin{array}
[c]{ccc}%
1 & 1 & 1
\end{array}
\right)  =\left(
\begin{array}
[c]{ccc}%
1 & 1 & 1\\
1 & 1 & 1\\
1 & 1 & 1
\end{array}
\right)  \ ,\\
P  &  =\left(
\begin{array}
[c]{ccc}%
1 & 0 & 0\\
0 & 0 & 0\\
0 & 0 & 0
\end{array}
\right)  =\frac{1}{64}\left(
\begin{array}
[c]{ccc}%
0 & 8 & 0\\
2 & 0 & -2\\
1 & -2 & 1
\end{array}
\right)  \left(
\begin{array}
[c]{ccc}%
1 & 1 & 1\\
1 & 1 & 1\\
1 & 1 & 1
\end{array}
\right)  \left(
\begin{array}
[c]{ccc}%
0 & 2 & 1\\
8 & 0 & -2\\
0 & -2 & 1
\end{array}
\right)  \ .
\end{align*}
For this example, eigenvalues and a particular choice of \emph{orthogonal}
eigenvectors for $B$ are given by%
\[
B\left(
\begin{array}
[c]{c}%
1\\
1\\
1
\end{array}
\right)  =3\left(
\begin{array}
[c]{c}%
1\\
1\\
1
\end{array}
\right)  \ ,\text{ \ \ and \ \ }B\left(
\begin{array}
[c]{c}%
2\\
-1\\
-1
\end{array}
\right)  =B\left(
\begin{array}
[c]{c}%
0\\
1\\
-1
\end{array}
\right)  =0\text{ .}%
\]
For spin $j=3/2$,
\begin{align*}
B  &  =\left(
\begin{array}
[c]{c}%
1\\
1\\
1\\
1
\end{array}
\right)  \left(
\begin{array}
[c]{cccc}%
1 & 1 & 1 & 1
\end{array}
\right)  =\left(
\begin{array}
[c]{cccc}%
1 & 1 & 1 & 1\\
1 & 1 & 1 & 1\\
1 & 1 & 1 & 1\\
1 & 1 & 1 & 1
\end{array}
\right)  \ ,\\
P  &  =\left(
\begin{array}
[c]{cccc}%
1 & 0 & 0 & 0\\
0 & 0 & 0 & 0\\
0 & 0 & 0 & 0\\
0 & 0 & 0 & 0
\end{array}
\right)  =\frac{1}{2304}\left(
\begin{array}
[c]{cccc}%
-3 & 27 & 27 & -3\\
-1 & 27 & -27 & 1\\
3 & -3 & -3 & 3\\
1 & -3 & 3 & -1
\end{array}
\right)  \left(
\begin{array}
[c]{cccc}%
1 & 1 & 1 & 1\\
1 & 1 & 1 & 1\\
1 & 1 & 1 & 1\\
1 & 1 & 1 & 1
\end{array}
\right)  \left(
\begin{array}
[c]{cccc}%
-3 & -1 & 3 & 1\\
27 & 27 & -3 & -3\\
27 & -27 & -3 & 3\\
-3 & 1 & 3 & -1
\end{array}
\right)  \ .
\end{align*}
Particular orthogonal eigenvectors and eigenvalues of $B$ are now given by%
\[
B\left(
\begin{array}
[c]{c}%
1\\
1\\
1\\
1
\end{array}
\right)  =4\left(
\begin{array}
[c]{c}%
1\\
1\\
1\\
1
\end{array}
\right)  \ ,\text{ \ \ and \ \ }B\left(
\begin{array}
[c]{c}%
1\\
-1\\
0\\
0
\end{array}
\right)  =B\left(
\begin{array}
[c]{c}%
0\\
0\\
1\\
-1
\end{array}
\right)  =B\left(
\begin{array}
[c]{c}%
1\\
1\\
-1\\
-1
\end{array}
\right)  =0\ .
\]
The generalization to other $j$ is straightforward.

\newpage

\subsection*{Appendix F: \ Trace of a matrix star product}

Let us pursue the business of writing the matrix biorthonormality in the main
text as a trace relation.%
\begin{equation}
\delta_{m,n}=\operatorname*{Trace}\left(  G~S^{n}~B~S^{m}\right)  \ ,
\tag{F1}\label{TraceNormCycled}%
\end{equation}
where, as before, $B$ is a singular matrix with all entries equal to $1$. \ We
have cycled the matrices in (\ref{TraceNorm}) into the ordering shown in
(\ref{TraceNormCycled}) to emphasize a possible analogy to distributions
integrated over phase space \cite{CTQMPS}. \ The second ordering suggests more
clearly which of $B$ \& $G$ corresponds to a \textquotedblleft
measure\textquotedblright\ (note that $G$ is \emph{non}singular)\ and which to
a \textquotedblleft star product\textquotedblright\ (once again, note that $B$
is singular). \ 

Moreover, for any spin $j$ the matrix $B$ is obtained by acting on the
projector
\begin{equation}
P=\left(
\begin{array}
[c]{cccc}%
1 & 0 & 0 & \cdots\\
0 & 0 & 0 & \cdots\\
0 & 0 & 0 & \cdots\\
\vdots & \vdots & \vdots & \ddots
\end{array}
\right)  \tag{F2}%
\end{equation}
with the Vandermonde matrix and its transpose: $\ $%
\begin{equation}
B=V\left[  j\right]  ~P~V\left[  j\right]  ^{\text{transpose}}=\left(
\begin{array}
[c]{cccc}%
1 & 1 & 1 & \cdots\\
1 & 1 & 1 & \cdots\\
1 & 1 & 1 & \cdots\\
\vdots & \vdots & \vdots & \ddots
\end{array}
\right)  \ . \tag{F3}\label{BAsConjugatedProjector}%
\end{equation}
Or conversely, $B$ is diagonalized, and rescaled by $1/\left(  2j+1\right)  $,
by acting with $V^{-1}$ and its transpose:%
\begin{equation}
P=V^{-1}\left[  j\right]  \mathbb{~}B~V^{-1}\left[  j\right]
^{\text{transpose}}\ . \tag{F4}\label{PAsConjugatedB}%
\end{equation}
Therefore, from $G=\left(  V^{-1}\right)  ^{\dag}\left(  V^{-1}\right)  $, the
orthonormality relation may be written as a trace-projection of similarity
transformed (i.e. conjugated) powers of $S$.%
\begin{equation}
\delta_{m,n}=\operatorname*{Trace}\left[  \left(  V^{-1}~S^{n}~V\right)
~P~\left(  V^{-1}~S^{m}~V\right)  ^{\dag}\right]  \ , \tag{F5}%
\label{TraceNorm2}%
\end{equation}
where, in the chosen basis that diagonalizes $S$, there is no distinction
between hermitian conjugation and taking the transpose since all matrices are
real. \ However, having written it in this form, the relation
(\ref{TraceNorm2}) is valid in \emph{any} basis upon using suitably
transformed $V$ and $P$. \ Note that $P$\ will remain a projector in any other
basis obtained by a similarity transformation. $\ $Also, $P$\ will be
hermitian and positive-semi-definite in any other basis obtained by a unitary
transformation. \ 

Compare the result in (\ref{TraceNorm2})\ with the usual trace-norm of an
arbitrary matrix $M$, namely, $\operatorname*{Trace}\left(  MM^{\dag}\right)
$. \ As has been shown here, orthonormality of the powers of $V^{-1}SV$
results from a modified product within the trace, a modification
produced\ simply by inserting $P$, albeit a singular matrix. \ As is
well-known, associativity is not destroyed by such insertions to produce
modified products of three or more matrices, hence it may be useful to think
of the modification as a \emph{matrix star product}, with $M_{1}\star
M_{2}\equiv M_{1}PM_{2}$. \ The fact that $P$\ is singular should not be cause
for much concern, at least for the problem at hand, since already $S$ is not
invertible for integer $j$.

As previously suggested, these trace-norm and modified product results are
somewhat analogous to those for quantum distribution functions on phase space
\cite{CTQMPS}. \ A well-known result for discretely indexed pure-state Wigner
functions, $\operatorname*{W}_{a}\left(  x,p\right)  $, is that $2\pi\hbar
\int\operatorname*{W}_{a}\left(  x,p\right)  \star\operatorname*{W}_{b}\left(
x,p\right)  dxdp=\delta_{a,b}$ where the star product is that discovered by
Groenewold, namely, $\star=\exp\frac{i\hbar}{2}\left(  \overleftarrow{\partial
}_{x}\overrightarrow{\partial}_{p}-\overleftarrow{\partial}_{p}%
\overrightarrow{\partial}_{x}\right)  $. \ In this case the Wigner functions
are actually their own duals in the sense that the modified product may be
eliminated by phase-space integrations by parts without changing the result
(assuming there are no net contributions from the boundaries of phase space).
\ Thus$\ 2\pi\hbar\int\operatorname*{W}_{a}\left(  x,p\right)
\operatorname*{W}_{b}\left(  x,p\right)  dxdp=\delta_{a,b}$. \ Moreover, in
this case the modified product itself reduces to $2\pi\hbar\operatorname*{W}%
_{a}\left(  x,p\right)  \star\operatorname*{W}_{b}\left(  x,p\right)
=\operatorname*{W}_{a}\left(  x,p\right)  ~\delta_{a,b}$ with the individual
Wigner functions normalized to $\int\operatorname*{W}_{a}\left(  x,p\right)
dxdp=1$. \ But none of these last three statements apply by analogy to powers
of spin matrices, their modified products, or their traces, at least not as
described above. \ 

Other analogies can be found for other phase space distributions \cite{Lee}.
\ For example, Husimi distributions, $\operatorname*{H}_{a}\left(  x,p\right)
$, also obey $2\pi\hbar\int\operatorname*{H}_{a}\left(  x,p\right)
\ast\operatorname*{H}_{b}\left(  x,p\right)  dxdp=\delta_{a,b}$ with an
appropriately modified $\ast$ product, but in this case the functions are
\emph{not} their own duals because integrations by parts do not eliminate the
$\ast$. \ Rather, the dual functions are given by Glauber-Sudarshan
distributions, $\operatorname*{GS}_{a}$, such that $2\pi\hbar\int%
\operatorname*{GS}_{a}\left(  x,p\right)  \operatorname*{H}_{b}\left(
x,p\right)  dxdp=\delta_{a,b}$, a result that is\ more closely analogous to
the matrix statement (\ref{OrthoTrace}).\newpage

\subsection*{Appendix G: \ Periodicized monomial approximations}

The coefficients $A_{0,1,2,3,4,5}^{\left[  j\right]  }\left(  \theta\right)  $
are plotted here for $j=69$ in blue, and for $j=137/2$ in red. \ The slope of
each red (fermionic) curve plotted on the right is given exactly by the red
curve plotted to its immediate left. \ Similarly, the slope of any blue
(bosonic) curve on the left is given by the blue curve to its right, but in
the row above.

\noindent\hspace{-0.25in}%
{\parbox[b]{3.6588in}{\begin{center}
\includegraphics[
height=2.2707in,
width=3.6588in
]%
{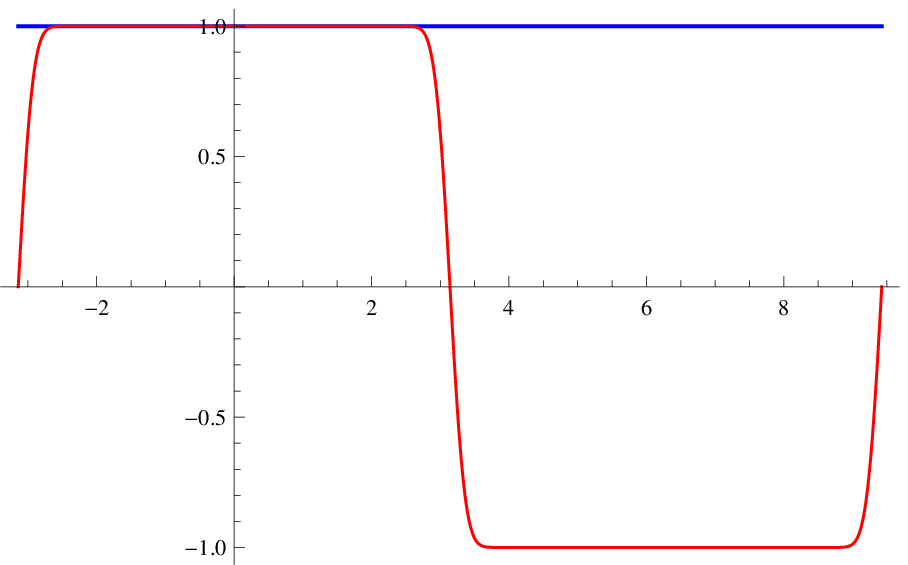}%
\\
$A_{0}^{\left(  j\right)  }\left(  \theta\right)  $ versus $\theta$%
\end{center}}}
\ \ \
{\parbox[b]{3.6588in}{\begin{center}
\includegraphics[
height=2.2707in,
width=3.6588in
]%
{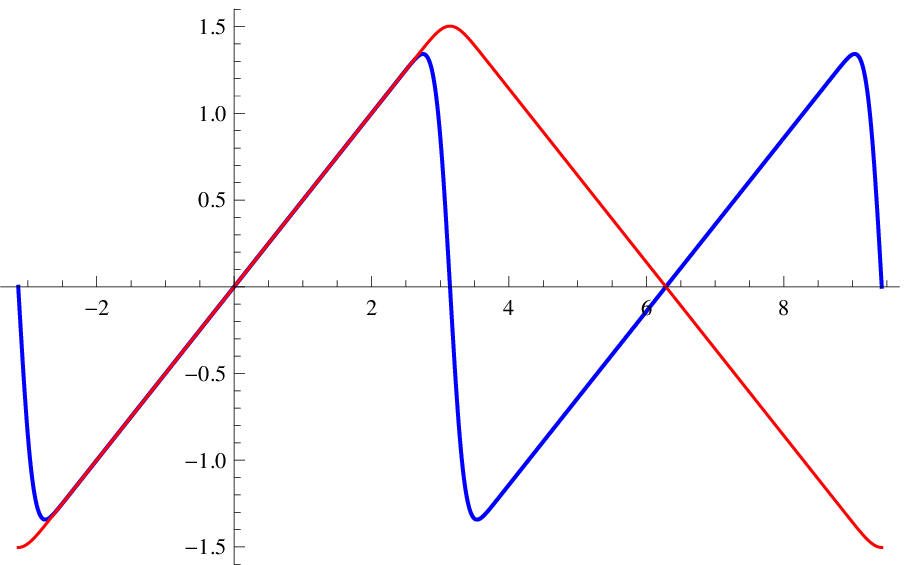}%
\\
$A_{1}^{\left(  j\right)  }\left(  \theta\right)  $ versus $\theta$%
\end{center}}}

\noindent\hspace{-0.25in}%
{\parbox[b]{3.6588in}{\begin{center}
\includegraphics[
height=2.2707in,
width=3.6588in
]%
{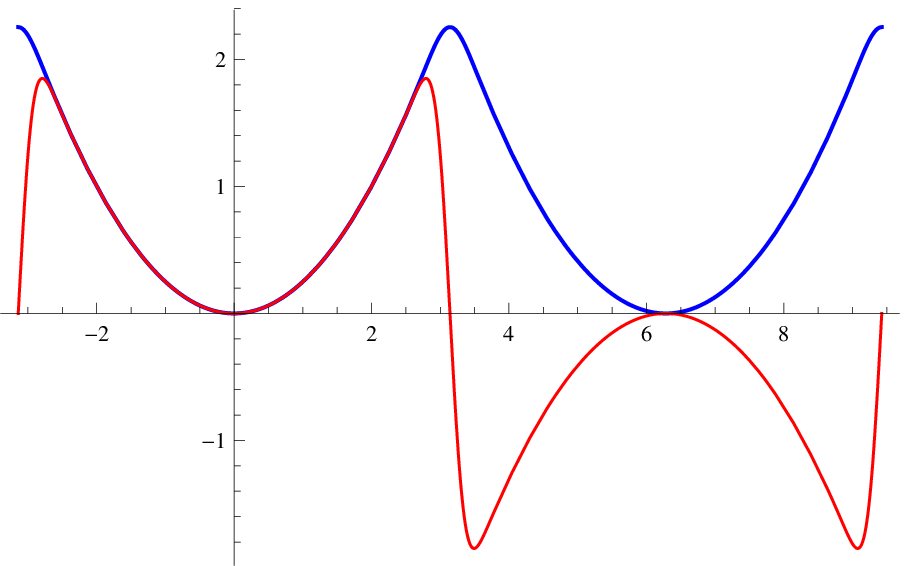}%
\\
$A_{2}^{\left(  j\right)  }\left(  \theta\right)  $ versus $\theta$%
\end{center}}}
\ \ \
{\parbox[b]{3.6588in}{\begin{center}
\includegraphics[
height=2.2707in,
width=3.6588in
]%
{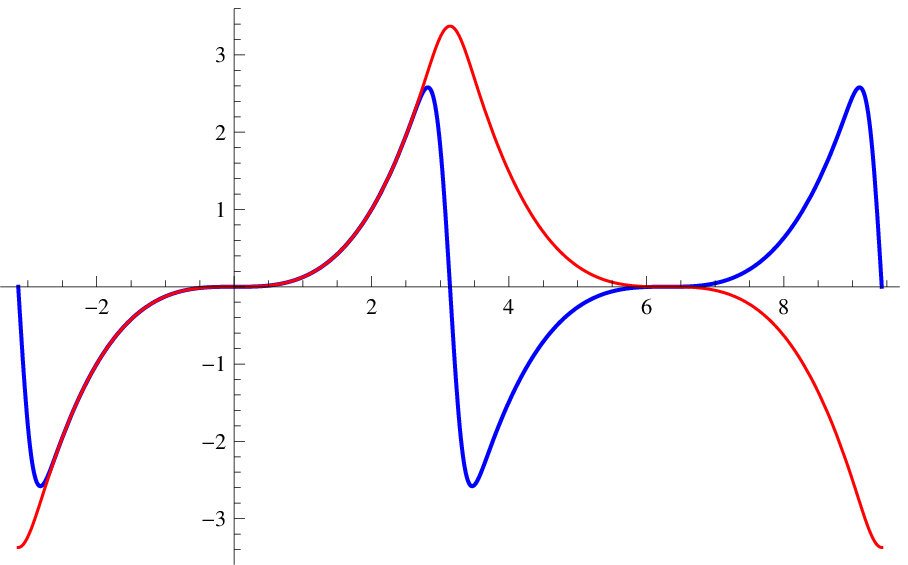}%
\\
$A_{3}^{\left(  j\right)  }\left(  \theta\right)  $ versus $\theta$%
\end{center}}}

\noindent\hspace{-0.25in}%
{\parbox[b]{3.6588in}{\begin{center}
\includegraphics[
height=2.2707in,
width=3.6588in
]%
{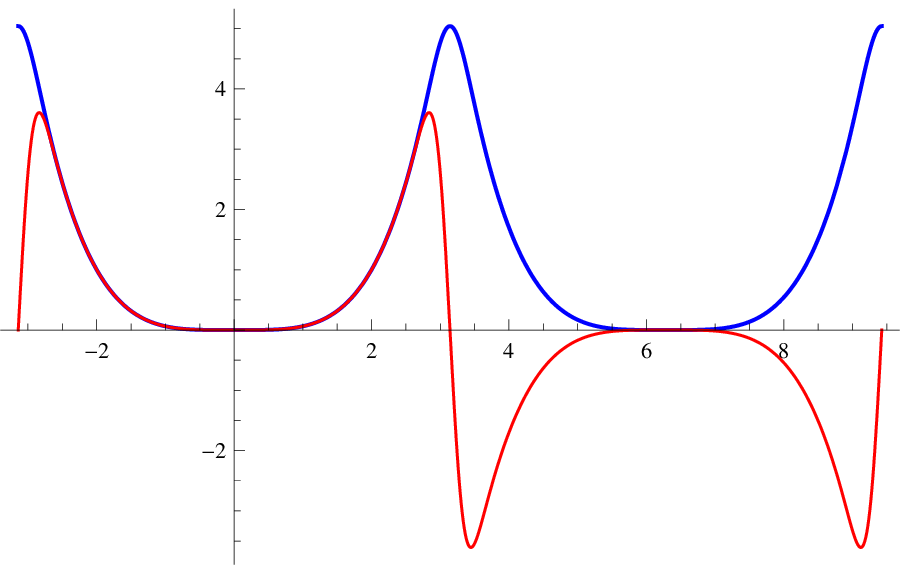}%
\\
$A_{4}^{\left(  j\right)  }\left(  \theta\right)  $ versus $\theta$%
\end{center}}}
\ \ \
{\parbox[b]{3.6588in}{\begin{center}
\includegraphics[
height=2.2707in,
width=3.6588in
]%
{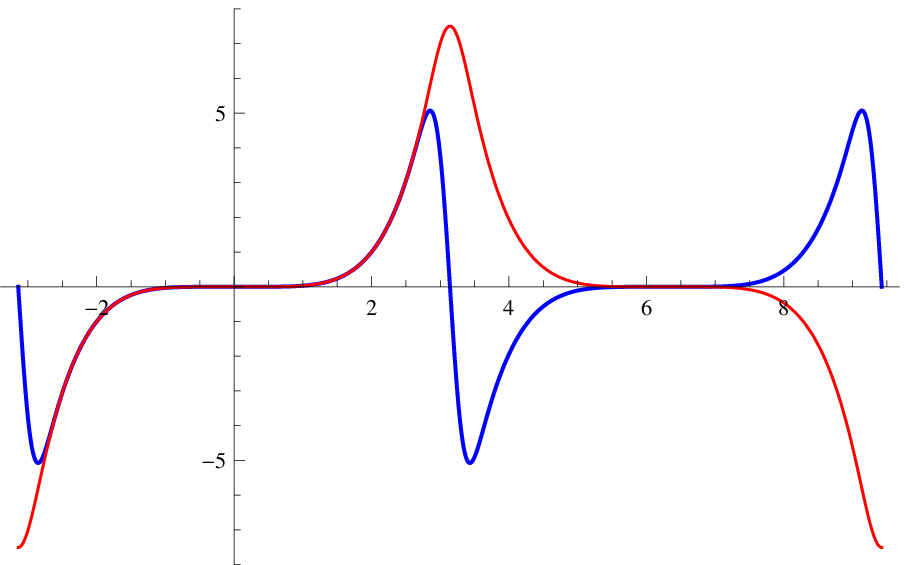}%
\\
$A_{5}^{\left(  j\right)  }\left(  \theta\right)  $ versus $\theta$%
\end{center}}}

\end{document}